\documentclass[aps,pra,floatfix,twocolumn,showpacs,10pt,longbibliography]{revtex4-2}
\usepackage{amsmath}
\usepackage{braket} 
\usepackage{url}
\usepackage[ruled,vlined]{algorithm2e}
\usepackage{algorithmic}
\usepackage{graphicx}
\usepackage{dcolumn}
\usepackage{bm}
\usepackage{amssymb}
\usepackage{rotating}
\usepackage[abs]{overpic}
\usepackage{xcolor}
\usepackage{booktabs}
\usepackage{tikz}
\usepackage{tabularx}
\usepackage{ragged2e}
\usepackage{hyperref}
\hypersetup{colorlinks = true, citebordercolor={blue}, linkcolor={blue}, citecolor={blue}, urlcolor={blue}}

\DeclareUnicodeCharacter{2212}{-}
\DeclareUnicodeCharacter{2009}{-}

\begin{document}



\title{Enhanced Shadow Tomography of Molecular Excited States from Enforcing $N$-representability Conditions by Semidefinite Programming}
\author{Irma Avdic and David A. Mazziotti}

\email{damazz@uchicago.edu}

\affiliation{Department of Chemistry and The James Franck Institute, The University of Chicago, Chicago, IL 60637 USA}

\date{Submitted July 20, 2024}

\begin{abstract}
Excited-state properties of highly correlated systems are key to understanding photosynthesis, luminescence, and the development of novel optical materials, but accurately capturing their interactions is computationally costly. We present an algorithm that combines classical shadow tomography with physical constraints on the two-electron reduced density matrix (2-RDM) to treat excited states. The method reduces the number of measurements of the many-electron 2-RDM on quantum computers by (\textit{i}) approximating the quantum state through a random sampling technique called shadow tomography and (\textit{ii}) ensuring that the 2-RDM represents an $N$-electron system through imposing $N$-representability constraints by semidefinite programming. This generalizes recent work on the $N$-representability-enhanced shadow tomography of ground-state 2-RDMs. We compute excited-state energies and 2-RDMs of the H$_4$ chain and analyze the critical points along the photoexcited reaction pathway from \textit{gauche}-1,3-butadiene to bicyclobutane via a conical intersection. The results show that the generalized shadow tomography retains critical multireference correlation effects while significantly reducing the number of required measurements, offering a promising avenue for the efficient treatment of electronically excited states on quantum devices.

\end{abstract}

\maketitle


\section{Introduction}
Most fundamental life processes rely on photochemistry. Electronically excited states play a central role in the mechanism of light-induced reactions driving many important phenomena such as photosynthesis~\cite{Cerullo2002, Cheng_2009}, vision~\cite{Herbst2002}, and the protection of DNA from damage~\cite{Schultz2004, Yam2023, Hsu2017, Lindh2020}. However, excited-state computations are typically more difficult to compute than their ground-state counterparts. Ensuring orthogonality of the excited states to the lower-lying ones~\cite{Bustard1970, Glushkov2007, Tassi2013, Szabo1996} and accounting for the many determinants contributing to the excited-state wave function, i.e., the multireference character~\cite{Balkov1994, Parac2002, Mazziotti2003, Sundstrom2014}, represent just some of the challenges.  Using quantum computers offers an efficient paradigm for treating excited states because their quantum states can be directly prepared as a superposition of qubits without explicit classical representation or storage of the wave function~\cite{Head-Marsden2020, Bharti2022}.  Nonetheless, a significant part of realizing this potential depends upon achieving efficient, noise-resilient tomography of critically important observables (e.g., from the 1- and 2-electron reduced density matrices) with a minimal number of measurements.



While full quantum state tomography is possible, it is highly inefficient, requiring an exponential number of parameters to describe an $N$-particle state as well as an exponentially large amount of post-processing~\cite{ODonnell2016}. Various approaches to quantum state tomography of generic, complex quantum systems have been developed in recent years~\cite{Lanyon2017, Yen2020, Reagor2018, Cramer2010, Rambach2021, Haah2017,Gupta2021, Gupta2022, Gard2020, Tang2021}, including those based on the theory of classical shadows~\cite{Aaronson2020, Huang2020, Huang2022, Hu2023, McGinley2022, OGorman2022, Low2022, Fawzi.2024}. By focusing only on the physically relevant quantities ``hidden'' in the reduced-dimensional space of the system, the classical shadow theory allows for learning of many properties of an $N$-dimensional quantum system from partial information~\cite{Zhao2021, Peng2023, Hearth.2023, Gyurik.2023, Jerbi.2023, Struchalin.2021, Lewis.2024, O'Gorman.2022, Koh.2022, Ippoliti.2023, Nguyen.2022, Helsen.2023ocb, Coopmans.2023, Guzman.2023, Jnane.2024, Wu.2023, Akhtar.2023, Hu.2024, Truger.2024, Caprotti.2024, Majsak.2024, Levy.2024, Bu.2024, Becker.2024}. While the classical shadow approach is able to predict many observables efficiently, it does not guarantee that the sampled outcomes will correspond to physical states, especially on noisy intermediate-scale quantum (NISQ) devices.  This challenge is only amplified in the case of excited states where it is compounded by issues like state orthogonality and strong correlation, arising from the energetic degeneracies of the molecular orbitals.


Here, we combine the concept of shadow tomography and reduced density matrix theory to accelerate the convergence of classical shadow tomography for excited states. In reduced density matrix theory, the energy of a system is variationally minimized as a functional of the two-electron reduced density matrix (2-RDM). The energy of any stationary state is a linear functional of the 2-RDM~\cite{Coleman1963, Lowdin1955, Meyer1955, Coleman2000}, but the 2-RDM must be restricted by additional, nontrivial constraints, known as $N$-representability conditions~\cite{Coleman1963, Garrod1969, Kummer1967, Erdahl1978, Mazziotti2012, Mazziotti2012_2, Mazziotti2004, Mazziotti2016, mazziotti2001, Gidofalvi2005, gidofalvi_mazziotti_2006, Mazziotti2006_2, Mazziotti2023}, to ensure that it is representable by at least one $N$-electron density matrix. In the context of tomography, the $N$-representability conditions provide a tremendous advantage because they significantly constrain the convex set of 2-RDMs, thereby decreasing the degrees of freedom to be sampled by shadow tomography. Here we impose the $N$-representability conditions through a novel post-processing algorithm on a classical device, in which the 2-RDM is extracted from a very small number of shadows by semidefinite programming (SDP)~\cite{VB1996, M2004,M2011}. The fusion of SDP with minimal shadows on quantum devices provides the framework for more efficient and more accurate quantum molecular computations of excited states on quantum devices. This algorithm generalizes the previous work on ground-state shadow-enhanced tomography of the 2-RDM~\cite{Avdic2024}.

We first present the $N$-representability-enhanced shadow tomography and its generalization to excited states. We then demonstrate the approach by studying the first four low-lying singlet excited states of the H$_4$ chain and the photoexcited pathway from \textit{gauche}-1,3-butadiene to bicyclobutane via a conical intersection. We show that the excited-state shadow tomography with $N$-representability conditions is able to capture single- and multireference correlation effects and achieve chemical accuracy ($\approx$ 1 kcal/mol) with reasonable computational resources. Moreover, we illustrate how considering $N$-representability in the context of shadow tomography provides a significant advantage in reaching high accuracy from fewer measurements even in highly multireference systems such as conical intersections. Although the presented tomography is demonstrated through classical calculations here, it has important applications to accelerating and enhancing the quantum tomography of excited states on quantum devices. Not only does the approach accelerate the convergence of the classical shadows towards the 2-RDM but it also provides a built-in error mitigation for correcting preparation and measurement errors in the 2-RDM with respect to known physical constraints---the $N$-representability conditions.



\section{Theory}
The two-electron reduced density matrix (2-RDM) is defined as,
\begin{equation}
    ^2D^{ij}_{kl} = \bra{\Psi}\hat{a}^\dagger_{i}\hat{a}^\dagger_{j}\hat{a}^{}_{l}\hat{a}^{}_{k}\ket{\Psi},
\end{equation}
where $\hat{a}_{i}^{\dagger}$ creates an electron in orbital $i$ and $\hat{a}^{}_{i}$ annihilates an electron in orbital $i$. To minimize the energy as a functional of the 2-RDM, $E[^2D]$ while making sure the particle-particle ($^2D$) particle-hole ($^2G$) and the hole-hole ($^2Q$) matrices are positive semidefinite and related through linear mappings, we use semidefinite programming (SDP). The following is a description of such a program for a variational 2-RDM calculation,
\begin{align}
    \min_{^2D \in {}^N_2\!\Tilde{P}}& \hspace{0.2cm} E[^2D] \label{eq:E} \\
\text{such that}
    \hspace{0.2cm} ^2D &\succeq 0 \\
                   ^2Q &\succeq 0 \\
                   ^2G &\succeq 0 \\
                   {\rm Tr}(^2D) &= N(N-1) \\
                   ^2Q &= f_Q(^2D) \\
                   ^2G &= f_G(^2D). \label{eq:sdp-regular}
\end{align}
Here, $^{2} Q$ and $^{2} G$ are the hole-hole and particle-hole matrices whose elements are given by
\begin{align}
^{2} Q^{kl}_{ij} & = \langle \Psi | {\hat a}^{}_{k} {\hat a}^{}_{l} {\hat a}^{\dagger}_{j} {\hat a}^{\dagger}_{i}| \Psi \rangle
\label{eq:Q} \\
^{2} G^{il}_{kj} & = \langle \Psi | {\hat a}^{\dagger}_{i} {\hat a}^{}_{l} {\hat a}^{\dagger}_{j} {\hat a}^{}_{k}| \Psi \rangle . \label{eq:G}
\end{align}
The positivity constraint $M \succeq 0$ enforces the matrix $M$ to be positive semidefinite, and ${\rm Tr}(^2D) = N(N-1)$ ensures the correct normalization. The functions $f_Q$ and $f_G$ represent the linear mappings between $^{2}Q$ and $^{2} D$ and $^{2} G$ and $^{2} D$, respectively, derived from the rearrangement of creation and annihilation operators in Eqs.~(\ref{eq:Q}) and~(\ref{eq:G})~\cite{M2007}. ${}^N_2\!\Tilde{P}$ is a convex set of approximately $N$-representable 2-RDMs.

Building upon our previous work for ground states~\cite{Avdic2024} as well as work by Huang et al.~\cite{Huang2020} and Aaronson~\cite{Aaronson2020}, we introduce an excited-state shadow tomography that combines the reduced density matrix theory above with classical shadows. We create a collection of two-electron classical shadows $S_{n}$ of a quantum state $\ket{\Psi}$, indexed by the integer $n$, by applying a set of unitary transformations $\hat{U}_{n}$ and then measuring the resulting states in the computational basis
\begin{equation}
    S_{n}^{pq} = \bra{\Psi}\hat{U}^{\dagger}_{n}\hat{a}^\dagger_{p}\hat{a}^\dagger_{q}\hat{a}^{}_{q}\hat{a}^{}_{p}\hat{U}_n\ket{\Psi},
    \label{eq:shadow}
\end{equation}
where $\hat{U}_{n} = \exp{({\sum_{uv}A^{uv}_{n}\hat{a}^{\dagger}_{u}\hat{a}^{}_{v}})}$, with $A_n$ being a one-body anti-Hermitian matrix, and $\ket{\Psi}$ a ground- or excited-state wave function. Each shadow $S_{n}^{pq}$ measures the diagonal elements of the 2-RDM after the $n^{\text{th}}$ one-body unitary transformation. The transformations, which rotate the orbitals into randomly selected bases, are sampled using the Haar measure~\cite{Haar1933}.

The collection of shadows forms a system of equations whose solution determines the matrix elements of the 2-RDM
\begin{equation}
    \label{eq:Spq}
    S_{n}^{pq} = \sum_{ijkl}{U_{n}^{pi} U_{n}^{pj} \, {^2D^{ij}_{kl}} \, U_{n}^{ql} U_{n}^{qk}},
\end{equation}
with $U_{n} = \exp{(A_{n})}$. Rather than solving this system directly, however, we add this system of equations to the semidefinite program for the variational 2-RDM calculation, given above.  The solution of the shadow constraints by SDP allows us to find the solution of the underdetermined shadow constraints that minimizes the energy and satisfies the $N$-representability conditions~\cite{Avdic2024}. From this perspective, the SDP and $N$-representability conditions serve to regularize the solution of the shadow tomography, accelerating its convergence towards an accurate, physically meaningful solution.  Conversely, we can also view the shadow constraints as modifying the variational 2-RDM method to restrict its domain of the minimization to be consistent with the prior information obtained from the state measurements on the quantum device.

A sufficiently large set of these shadows, $S_n$ allows full prediction of the 2-RDM or its expectation values without a full tomography cost of the $N$-electron density matrix or even the 2-RDM since only diagonal parts of the 2-RDM are being measured. Moreover, because the diagonal parts of the 2-RDM commute, they can be measured simultaneously (i.e., in parallel) on the quantum device.  The most efficient procedure constructs just enough shadows to approximate the 2-RDM, subject to an approximate set of necessary $N$-representability conditions, such that its properties (e.g., energy, dipole, or orbital occupations) are within a target range.  Importantly, each of the linear equations in Eq.~(\ref{eq:Spq}), relating the measured shadows to the underlying 2-RDM, can be relaxed to a pair of inequalities that account for noise in the shadows through error parameters, reflecting the level of noise (refer to Ref.~\cite{Avdic2024} for more details).

While the above procedure for excited states is similar to that previously introduced for ground states~\cite{Avdic2024}, the excited states have the additional challenge that they must converge to a solution that is orthogonal to the ground state and the lower-lying excited states. Solution of the SDP in Eqs.~(\ref{eq:E}-\ref{eq:sdp-regular}) yields a lower-bound approximation to the ground-state energy. Adding the classical shadows of the 2-RDM as constraints to the SDP causes the energy to converge from below to the energy of the quantum state being prepared on the quantum computer. For example, if the prepared state is an excited state computed by full configuration interaction (FCI), the addition of shadow constraints causes the energy to converge with the number of shadows to the FCI energy of the state from below. Importantly, the orthogonality of the quantum state is not added to the variational 2-RDM calculation but rather is enforced implicitly through the shadow constraints. The shadow constraints indirectly contain information about the orthogonality conditions through their description of the excited state. Consequently, the procedure can be applied to a single excited state without reference in the shadow tomography to the other states.



The $N$-representable shadow tomography is compatible with any quantum algorithm for excited states and is robust to statistical error in the 2-RDM such as in the presence of noise on quantum devices, as demonstrated in ~\cite{Avdic2024}. For example, the presented tomography can be combined with variational approaches for excited states such as the variational quantum deflation (VQD) and variational subspace expansion (VSE) algorithms~\cite{Higgott2019, Nakanishi2019} or non-variational approaches like the contracted quantum eigensolver (CQE)~\cite{Smart2021_2, Wang2023, Smart2024} algorithms that solve the contracted Schr{\"o}dinger equation (CSE). In each case, the enhanced shadow tomography may be used for more efficient and more accurate measurements of excited-state 2-RDMs and their observables.


\section{Applications}
To illustrate the effectiveness of the excited-state shadow variational 2-RDM (sv2RDM) method, we explore the algorithm's performance in two applications, the four lowest-lying singlet excited states of the strongly correlated linear H$_4$ chain~\cite{Suhai1994} and the photoexcited reaction pathway from \textit{gauche}-1,3-butadiene to bicyclobutane via a conical intersection~\cite{Sicilia2007, Snyder2011, Mazziotti2008, Boyn2022, Shevlin1988, Nguyen1995, Kinal2007, Lutz2008}. For the H$_4$ chain we generate the classical shadows from the full configuration interaction (FCI)~\cite{Craig1950} wave function with the minimal Slater-type-orbital (STO-3G) basis set~\cite{Hehre1969} and for the \textit{gauche}-1,3-butadiene to bicyclobutane pathway we use those from complete active space configuration interaction (CASCI)~\cite{Knowles1984} with a 10~electrons-in-10~orbitals (10,10) active space and the split-valence double-zeta polarized (6-31G*) basis set~\cite{Rassolov1998}. Using multiple excited states showcases the applicability of the SDP framework for accurately describing multi-reference, excited-state energies. All computations are performed using the Quantum Chemistry Package~\cite{rdmchem_2023} in Maple~\cite{maple_2023}.

\subsection{Low-lying H$_4$ excited states}
\begin{figure}[tb!]
    \centering
    \includegraphics[width=\hsize]{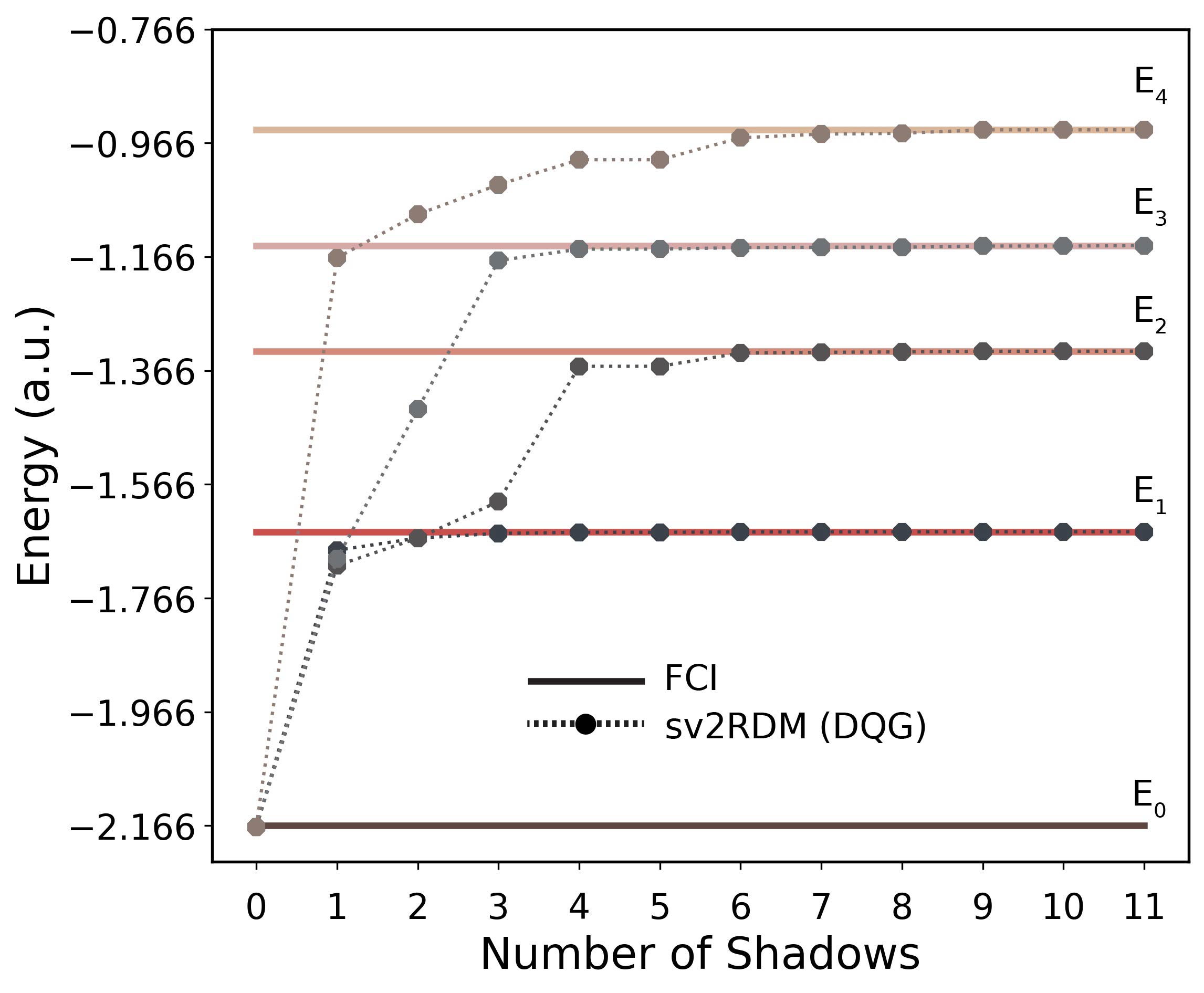}
    \caption{The energies (a.u.) of the four lowest-lying singlet excited states of the H$_4$ chain with 1.00 \r{A} H-H bonds are shown as a function of the number of shadows included in the sv2RDM. In each case, the exact convergence of energy to the FCI solution is reached within 11 shadows when all 2-positivity conditions (D, Q, and G) are included.}
    \label{fig:H4-excited-states}
\end{figure}

Figure~\ref{fig:H4-excited-states} shows excited-state energies of H$_4$ from the sv2RDM method with 2-positivity (DQG) conditions converging quickly with the number of shadows to those from the FCI solution. For the first excited state, adding a single classical shadow with D, Q, and G conditions to a shadowless variational 2-RDM calculation increases the convergence accuracy to the target state by 16 times. Similarly, in the calculation of higher excited states, the difference between the energy without shadows and the energy with one shadow shows the largest change---larger than the change from any subsequent additions of a single shadow.

The $N$-representability in the ground-state shadow tomography enhances the accuracy of observables with fewer measurements. In this work, the combination of shadow and $N$-representability constraints not only decreases the number of measurements relative to conventional shadow tomography for the 2-RDM but also imposes orthogonality of the targeted excited state relative to the ground state and the lower-lying excited states. While we focus on applying the technique to computing singlet states, the algorithm can treat any spin symmetry.

\begin{figure*}[t!]
    \centering
    \includegraphics[width=0.8\textwidth]{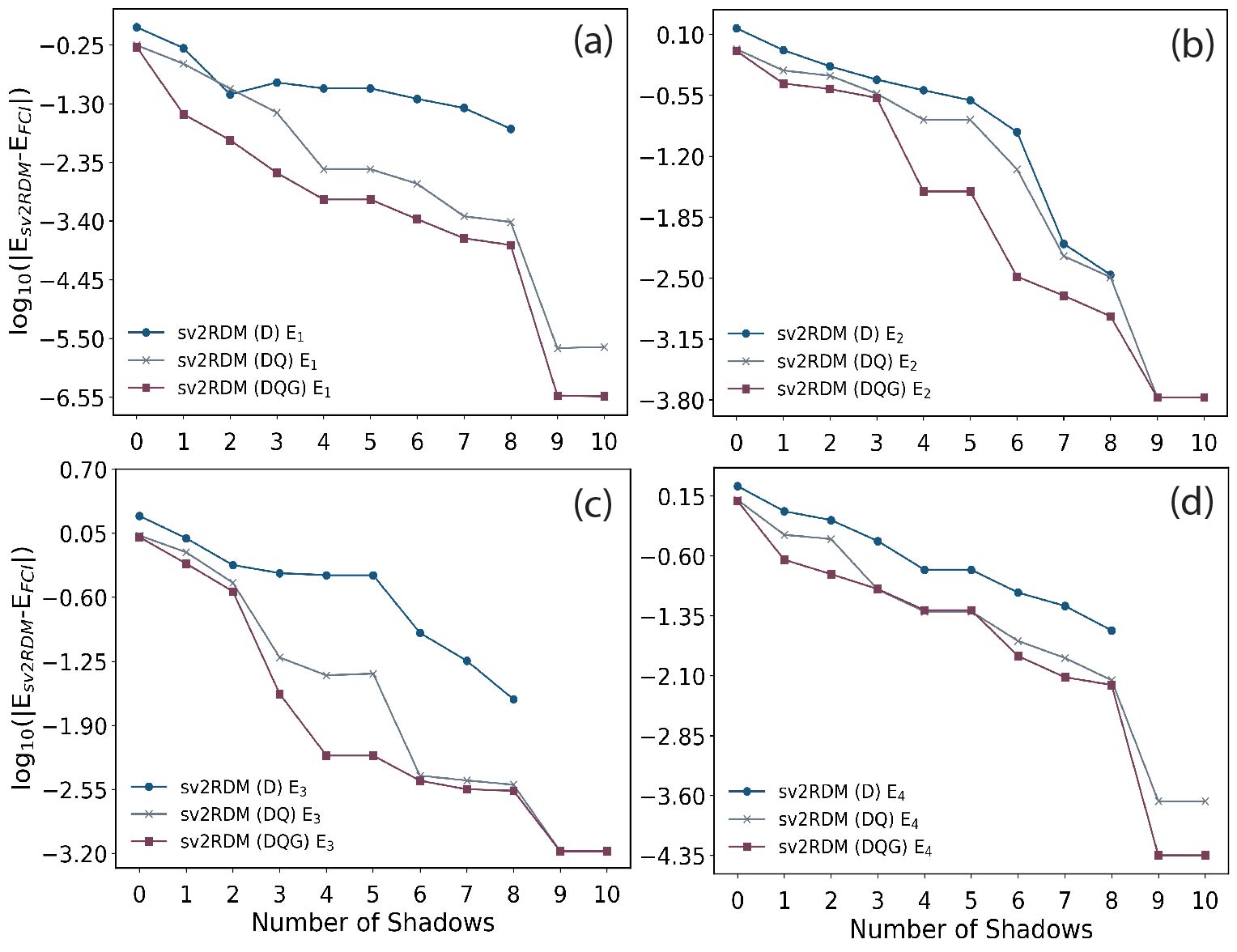}
    \caption{The absolute energy error of each of the four lowest-lying singlet excited states of H$_4$ for the different positivity conditions, plotted on the log$_{10}$ scale as a function of the number of shadows. Here, the panels show (a) the first excited state, E$_1$, (b) the second excited state, E$_2$, (c) the third excited state, E$_3$, and (d) the fourth excited state, E$_4$. The collective DQG conditions significantly improve the accuracy of any given excited-state calculation compared to the D condition alone.}
    \label{fig:H4-energy-error}
\end{figure*}

The logarithm of the absolute energy error for sv2RDM with different $N$-representability conditions is shown in Fig.~\ref{fig:H4-energy-error}(a)-(d). The energy errors for results with a single shadow decrease notably faster when all three $N$-representability conditions, rather than just the D condition, are considered. The sv2RDM (DQ) and (DQG) energies converge to the exact FCI energy with 11~generated shadows, while both reach chemical accuracy ($\approx$ 1 kcal/mol) in 7~shadows. Using only the D condition, we find that the sv2RDM calculation is within 17~mhartree of the target excited state, which also holds for the higher excited states.

\begin{table}[b!]
\caption{\label{H4_FCIerror}The normalized Frobenius norm of the 2-RDM relative to FCI in the STO-3G basis set is presented as a function of the number of shadows and $N$-representability conditions for the first four excited states of H$_4$ with 1.00 \r{A} H-H bonds. The DQG conditions significantly improve upon D for all shadows.}
\begin{ruledtabular}
\begin{tabular}{ccccc}
 State & Shadow(s) & D & DQ & DQG \\
  \hline
   E$_1$ & 1 & 0.37158371 & 0.28627519 & 0.09754087  \\
     & 5 & 0.07565879 & 0.00713184 & 0.00180852 \\
    & 9 & 0.03222087 & 0.00291267 & 0.00000549 \\
   E$_2$ & 1 & 0.42849229 & 0.20740997 & 0.21358258  \\
     & 5 & 0.14951900 & 0.12699556 & 0.02117810 \\
    & 9 & - & 0.00122819 &  0.00037389\\
   E$_3$ & 1 & 0.33982215 & 0.30487523 & 0.29051657  \\
     & 5 & 0.19731070 & 0.03114331 & 0.00379993 \\
    & 9 & - & 0.00625127 & 0.00045033 \\
   E$_4$ & 1 & 0.41517114 & 0.28399191 & 0.13804940  \\
     & 5 & 0.08075783 & 0.03443228 & 0.03624089 \\
    & 9 & - & 0.00418814 & 0.00178340
\end{tabular}
\end{ruledtabular}
\end{table}

Table~\ref{H4_FCIerror} illustrates the importance of including D, Q, and G conditions along with the shadow procedure for attaining the highest accuracy of the observables estimated from the 2-RDM. For example, with a single shadow and the DQG conditions, the Frobenius norm of the 2-RDM relative to FCI is approximately four times lower than that with only the D condition. With 9~shadows, the Frobenius norm of the difference between the sv2RDM (DQG) and FCI 2-RDMs is within the range of 10$^{-3}$ - 10$^{-6}$ for all studied excited states. Meanwhile, the excited-state energies computed with only the D condition show incomplete convergence to those from FCI with errors on the order of 10$^{-2}$. The results emphasize the necessity of including $N$-representability in the 2-RDM tomography for achieving both the target accuracy and the reduction in computational scaling with the number of shadows. The latter is especially true at the point of greatest difference between the initial and target states, as in the case of excited-state calculation from a ground-state reference.

Though the convergence pattern of the DQ and DQG curves in all panels of Fig.~\ref{fig:H4-energy-error} remains similar, it is important to note that in the higher excited states (panels (c) and (d)), including only the D and Q conditions may be enough for reaching chemical accuracy with the fewest computational resources. For example, in the case of the fourth excited state, 9 shadows are sufficient to reach convergence within chemical accuracy for both DQG and DQ conditions. Previously, we have demonstrated this interplay between the shadow behavior and $N$-representability conditions in the study of ground-state systems~\cite{Avdic2024}. Here, we demonstrate the efficiency of including even a subset of the conditions in the treatment of excited states with high accuracy.

\subsection{The photoexcited conversion of \textit{gauche}-1,3-butadiene to bicyclobutane}
The electrocyclic conversion of \textit{gauche} 1,3-butadiene (GBUT) to bicyclobutane (BIBUT) has been a subject of extensive study, with a special emphasis on the many conical intersections along its potential energy surface (PES)~\cite{Sicilia2007, Dick2008, Bernardi1997, Ostojic2001, Kinal2007, Lutz2008, Snyder2011, Mazziotti2008}. The conical intersections in these reactions offer an efficient radiationless deactivation or chemical transformation pathway of the reacting system. Previously, it has been shown that the transition from the photoexcited GBUT to BIBUT occurs via a disrotatory pathway~\cite{Snyder2011}, which is a conversion of a conjugated molecule to a cyclic structure where the ring closing of the two end carbons occurs by rotation in opposite directions according to the Woodward-Hoffmann rules~\cite{Woodward1970}. The disrotatory transition state (TSD) is a biradical with two approximately half-filled molecular orbitals and, as such, has a multireferenced wave function. We compute absolute energies of the critical points along the ground- and excited-state PES of GBUT and BIBUT via complete-active-space configuration-interaction (CASCI)~\cite{Knowles1984} method with a 10~electrons-in-10~orbitals (10,10) active space and in the split-valence double-zeta polarized (6-31G*) basis set~\cite{Rassolov1998}. Molecular geometries of GBUT, BIBUT, TSD, and the conical intersection (CI) are taken from Ref.~\cite{Sicilia2007} and are listed in the Supplemental Materials. All geometries were optimized using the complete-active-space self-consistent-field (CASSCF) method in the 6-31G* basis set with a (10,10) active space for GBUT, BIBUT, and TSD, and a (4,4) active space for the CI and state averaging with a weight of [1, 1].

\begin{figure}[h!]
    \centering
    \includegraphics[width=\hsize]{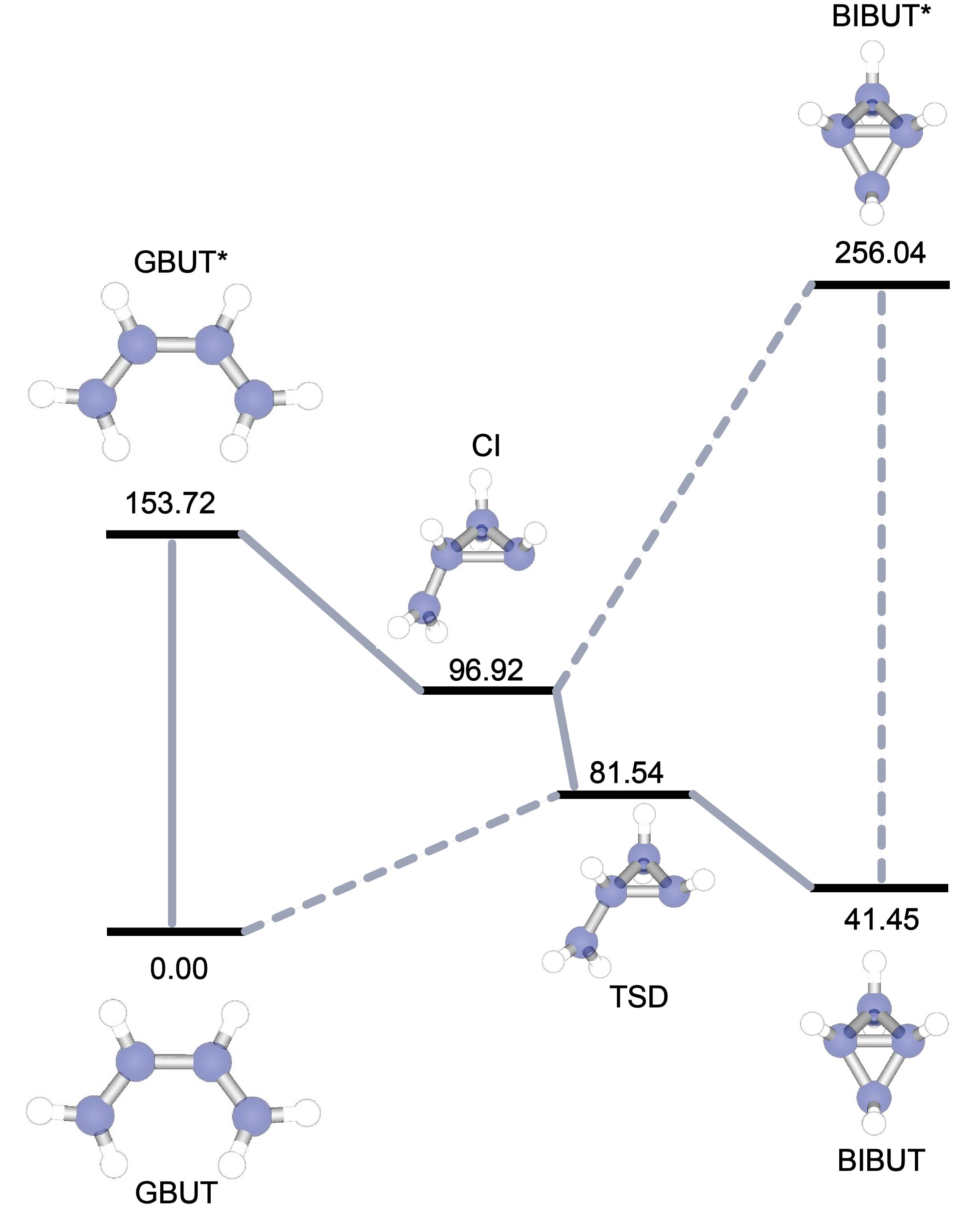}
    \caption{Critical points on the potential energy surface of the photoexcited conversion of \textit{gauche}-1,3-butadiene (GBUT) to bicyclobutane (BIBUT) computed via sv2RDM (DQG) in the 6-31G* basis set. The solid lines represent a path for the conversion of the excited-state reactants to ground-state products. The dashed lines represent the reverse pathway. The disrotatory transition state is labeled as TSD and the conical intersection is labeled as CI. Energies relative to GBUT are computed in kcal/mol.}
    \label{fig:pathway}
\end{figure}

The total and relative energies of the stationary points of the ground- and excited-state PES calculated using CASCI and sv2RDM (DQG) are reported in Table~\ref{tot_and_real_energies}. The results confirm the monotonically decreasing reaction pathway from the photoexcited state of \textit{gauche}- 1,3-butadiene (GBUT*) to BIBUT through the CI. The qualitative description of this pathway is presented in Fig.~\ref{fig:pathway} with included geometries of the studied stationary points and their energies relative to GBUT in kcal/mol. Both CASCI and sv2RDM confirm the previously reported features of this relaxation~\cite{Mazziotti2008, Snyder2011, Nguyen1995, Kinal2007} with a monotonically decreasing energy pathway through the conical intersection, which is geometrically very similar to TSD. Remarkably, sv2RDM, a method fundamentally based on the ground-state 2-RDM, is able to predict the correct excitation energies of both ground- and excited-state systems to within 1~mhartree for GBUT/GBUT* and 5~mhartree for the CI/CI* structures. Most importantly, the energy gaps between the ground and excited states of each system are conserved with the sv2RDM method. This suggests that the classical shadow, when sampling from an accurate excited-state wave function, does not implicitly alter the degree of correlation present in the system and uniformly retains the correlation effects. Additionally, with sufficient shadows, exact convergence is possible, though the optimal choice of the $N$-representability conditions and the number of shadows used will depend on the target accuracy.

\begin{table*}[t!]
\centering
\caption{\label{tot_and_real_energies} Total energies (a.u.) and relative energies (kcal/mol), with respect to the \textit{gauche}-1,3-butadiene (GBUT) molecule, are reported for the ground and first excited states of GBUT, bicyclobutane (BIBUT), disrotatory transition state (TSD), and the conical intersection (CI) in the 6-31G* basis set. The reported sv2RDM values are computed with D, Q, and G conditions.}
\begin{ruledtabular}
\begin{tabular}{lcccrrcccc}
\multicolumn{1}{c}{Molecule} & \multicolumn{3}{c}{Total energies (a.u.)} & \multicolumn{2}{c}{Relative energies (kcal/mol)} \\ \hline
                             & CASCI & sv2RDM & Shadows & CASCI & sv2RDM \\ \hline
GBUT     & -155.021 & -155.022 & 100 & 0.00 & 0.00 \\
GBUT*    & -154.776 & -154.777 & 110 & 154.08 &  153.72 \\
CI       & -154.863 & -154.868 & 90 & 99.03 & 96.92  \\
CI*      & -154.861 & -154.866 & 90 & 100.48 & 98.44 \\
TSD      & -154.890 & -154.892 & 120 & 82.13 & 81.54 \\
TSD*     & -154.795 & -154.797 & 120 & 142.02 & 141.17 \\
BIBUT    & -154.954 & -154.956 & 80 & 42.17 & 41.45 \\
BIBUT*   & -154.610 & -154.614 & 90 & 257.88 & 256.04 \\
\end{tabular}
\end{ruledtabular}
\end{table*}

\begin{table*}[t!]
\centering
\caption{\label{GBUT_BIBUT_occ}Natural occupation numbers and von Neumann entropies (vNEs) are reported for the ground- and first-excited states of the \textit{gauche}-1,3-butadiene (GBUT), bicyclobutane (BIBUT), disrotatory transition state (TSD), and the conical intersection (CI) in the 6-31G* basis set from the sv2RDM (DQG) method. The values agree exactly with those from the CASCI method in the same basis set.}
\begin{ruledtabular}
\begin{tabular}{ccccccccc}
\multicolumn{1}{c}{Orbital index} & GBUT    & GBUT*     & BIBUT    & BIBUT*  & TSD    & TSD*     & CI       & CI*      \\ \hline
14                                & 0.9598   & 0.8004   & 0.9864   & 0.9852  & 0.9835  & 0.9798   & 0.9774    & 0.9726      \\
15                                & 0.9373   & 0.5267   & 0.9799   & 0.4995  & 0.5511  & 0.9333   & 0.7675    & 0.8441      \\
16                                & 0.0646   & 0.4976   & 0.0213   & 0.4939  & 0.4496  & 0.0682   & 0.2349    & 0.1577      \\
17                                & 0.0387   & 0.1750   & 0.0123   & 0.0158  & 0.0192  & 0.0166   & 0.0240    & 0.0271      \\
vNE              & 0.4028  & 1.1685  & 0.1694 & 0.7753  & 0.7801 & 0.3355 & 0.6548 & 0.5592
\end{tabular}
\end{ruledtabular}
\end{table*}

To assess the degree of electron correlation and, more generally, the amount of quantum information present in the studied systems, we compute the natural occupation numbers and von Neumann entropies (vNEs) in Table~\ref{GBUT_BIBUT_occ}. The computed vNE is defined as,
\begin{equation}
    \text{vNE} = -\sum_{i}^{r} n_i\text{ln}(n_i),
\end{equation}
where $n_i$ is the occupation number of the $i$-th active orbital and $r$ is the number of active orbitals. Natural orbitals are defined as the eigenfunctions of the one-electron RDM (1-RDM) and their eigenvalues are the natural occupations (NOs). The exact agreement in NOs between CASCI and sv2RDM is reported in Table~\ref{GBUT_BIBUT_occ}. Without any electron correlation, the vNE is zero and increases with more fractional NOs between 0 and 1. The reported values correspond to a single spin block since all systems are studied in their singlet states, where the two spin blocks of the 1-RDM are identical. Nearly degenerate occupation of orbitals 15 and 16 in GIBUT* and BIBUT*, the highest- and lowest-occupied Hartree-Fock orbitals, respectively, reflects a significant amount of multireference (biradical) correlation in these systems, also depicted by vNE values significantly greater than zero. With respect to the number of shadows generated in the SDP, sv2RDM shows the highest agreement with the CASCI energy values for the GBUT* system. This demonstrates that the classical shadow procedure implemented with the DQG conditions has the ability to reconstruct a state of high multireference character, a result particularly relevant for chemically accurate simulation of physically relevant systems with high correlation on quantum devices.

\begin{figure}[h!]
    \centering
    \includegraphics[width=0.95\hsize]{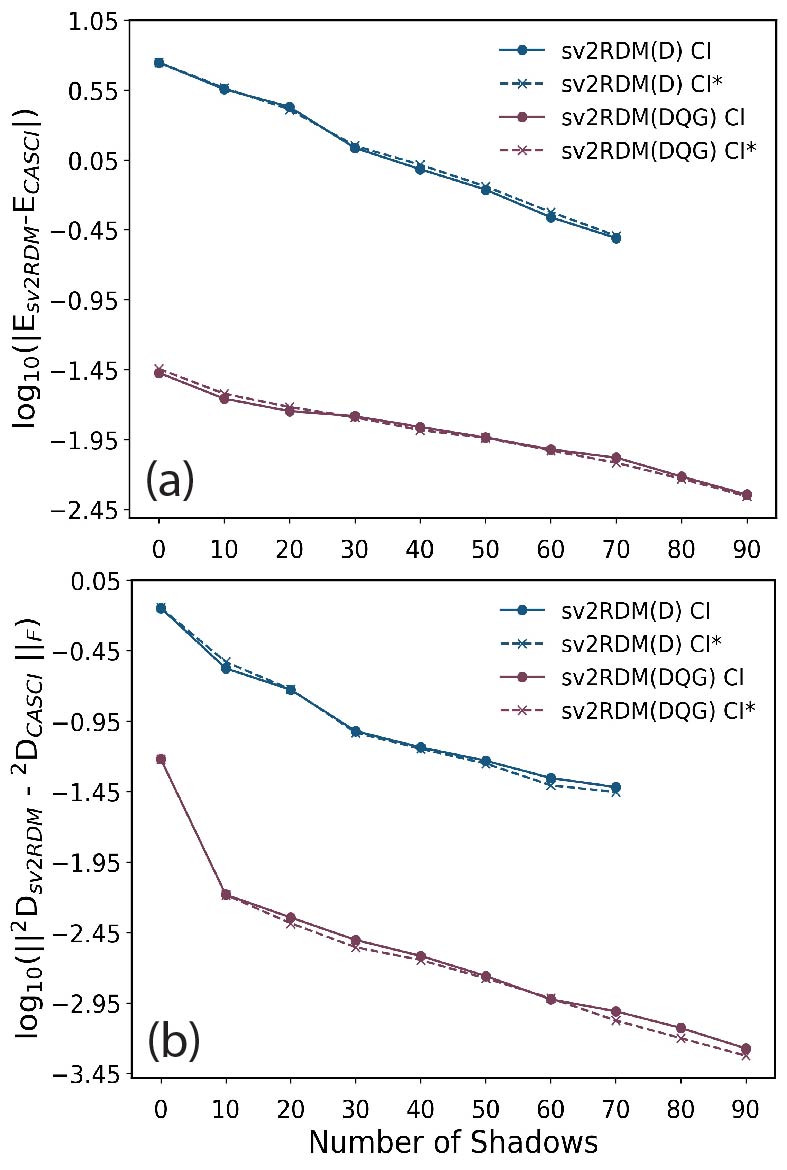}
    \caption{The absolute energy error between sv2RDM and CASCI (a) and the normalized Frobenius norm of the 2-RDM from sv2RDM relative to CASCI (b) are presented on the log$_{10}$ scale as functions of the number of shadows for the ground and excited states of the conical intersection (CI/CI*) with a comparison of calculations including only the D condition (blue lines) with those including D, Q, and G conditions (purple lines). It can be noticed the sv2RDM retains the small energy gap between the ground and excited states of the CI regardless of the number of positivity conditions included in the calculation, though the accuracy is significantly enhanced when all of the D, Q, and G  conditions are considered.}
    \label{fig:CI-errors}
\end{figure}

The conical intersection (CI) is a particularly interesting feature along this photoexcited conversion pathway, for it is a region where two potential energy surfaces intersect~\cite{Kuppel1984, Dick2008}. It is also a point of energetic degeneracy in the crossing of electronic states and thus, exhibits a significant amount of multireference correlation. The ground and excited states of the CI calculated with sv2RDM maintain a very narrow energy gap with one or all three of the positivity conditions, as shown in Fig.~\ref{fig:CI-errors}. Importantly, the accuracy of the converged values at each shadow step is much higher when all three of the 2-positivity conditions are considered. Hence, including the positivity conditions in the semidefinite procedure along with the shadow constraints plays a key role in efficiently and accurately reconstructing the highly correlated states. This conclusion is further supported by the results in Table~\ref{GBUTerror}, where the normalized Frobenius norm of the 2-RDM from the sv2RDM method relative to that from CASCI is reported. Evidently, including the Q and G conditions in addition to D suffices to provide near-to or chemically accurate results with the error in terms of the Frobenius norm being on the order of 10$^{-4}$ for all systems. Overall, the DQG conditions significantly improve upon D for any number of generated shadows.

\begin{table}[]
\caption{\label{GBUTerror}The normalized Frobenius norm of the 2-RDM relative to CASCI is reported as a function of the number of shadows and $N$-representability conditions for the ground and first excited states of \textit{gauche}-1,3-butadiene (GBUT), bicyclobutane (BIBUT), disrotatory transition state (TSD), and the conical intersection (CI) systems. In some cases, the sv2RDM with only the D condition was not able to converge beyond a certain number of shadows, which is illustrated by the absence of data in the table.}
\begin{ruledtabular}
\begin{tabular}{cccc}
 Molecule & Shadows & D & DQG\\
  \hline
   GBUT & 0 & 0.70721306 & 0.00561919  \\
   & 30 & 0.12277152 & 0.00198033 \\
     & 60 & 0.06623763 & 0.00079912 \\
     & 90 & - & 0.00033200 \\
    GBUT* & 0 & 0.70721307 & 0.11905626  \\
     & 30 & 0.12624018 & 0.00658575 \\
     & 60 & 0.05740913 & 0.00233879 \\
     & 90 & - & 0.00090734 \\
    BIBUT & 0 & 0.71642749 & 0.00517725  \\
    & 30 & 0.07324626 & 0.00108944 \\
     & 60 & 0.03608055 & 0.00040568 \\
     & 80 & - & 0.00020316 \\
    BIBUT* & 0 & 0.70949264  & 0.10192894  \\
    & 30 & 0.09922607 & 0.00278957 \\
     & 60 & 0.04188981 & 0.00122781 \\
     & 90 & - & 0.00047068 \\
    TSD & 0 & 0.70776552 & 0.00816901  \\
    & 30 & 0.10629265 & 0.00306653 \\
     & 60 & 0.05140730 & 0.00137910 \\
     & 90 & - & 0.00058986 \\
    TSD* & 0 & 0.71332641  &  0.11034467 \\
    & 30 & 0.08720601 & 0.00229741 \\
     & 60 & 0.04024479 & 0.00098947 \\
     & 90 & - & 0.00037923 \\
    CI & 0 & 0.71359496 & 0.06047515  \\
    & 30 & 0.09519725 & 0.00313828 \\
     & 60 & 0.04425476 & 0.00118907 \\
     & 90 & - & 0.00053367 \\
    CI* & 0 & 0.71523441 & 0.06047524  \\
    & 30 & 0.09311047 & 0.00278766 \\
     & 60 & 0.03932743 & 0.00121062 \\
     & 90 & - & 0.00047420 \\
\end{tabular}
\end{ruledtabular}
\end{table}

\section{Discussion and Conclusions}

In this work, we generalize the previously demonstrated $N$-representable shadow tomography of the 2-RDM~\cite{Avdic2024} to treat excited states. In the 2-RDM shadow tomography, the diagonal (classical) parts are measured with respect to randomly generated unitary transformations of the natural orbitals. The outputs of this procedure are classical shadows of an unknown quantum state that can be used to predict arbitrary expectation values. Our classical algorithm reconstructs the 2-RDM from a set of classical shadows while enforcing the necessary $N$-representability conditions to ensure the system of interest corresponds to an $N$-particle wave function. This procedure, we show here, can be generalized to treat excited states.  Significantly, the orthogonality of the excited states is implicitly contained in the classical shadows, allowing us to tomography various excited states without explicitly accounting for the orthogonality in the optimization. We can, therefore, apply the shadow tomography to any single excited state without knowledge of the ground state or the other excited states, making the algorithm applicable to any quantum algorithm for electronic excited states.

We demonstrated the effectiveness of $N$-representability conditions (DQG) in shadow tomography of the excited-state 2-RDM by calculating the four lowest-lying singlet excited states of the H$_4$ chain and the stationary points along the photoexcited conversion pathway of \textit{gauche}-1,3-butadiene to bicyclobutane via a conical intersection. The results illustrated accurate ground- and excited-state properties computed efficiently with approximately $n_sr^2$ measurements, where $n_s$ is the number of shadows and $r$ is the number of orbitals, with $n_s \ll r^2$ for target accuracy. Moreover, at this reduced computational cost, the algorithm preserved essential multireference correlation effects, which is especially important in studying physically relevant systems on quantum hardware.

Including the $N$-representability constraints in the shadow tomography of the reduced density matrices of many-body systems has the benefits of a reduced number of measurements, the ability to treat strong electron correlation, and a framework generalizable to calculating properties of larger many-fermion quantum systems. Here, our results showcase the significant advantage of merging $N$-representability constraints with shadow tomography, especially for excited states. Additionally, we demonstrate that including only a subset of the $N$-representability conditions, such as the D and Q conditions, can be highly effective for reaching target accuracy.

The present work lays the foundation for the application of the $N$-representable shadow tomography to calculating excited states with various quantum algorithms, including those based on the variational quantum eigensolver (VQE) as well as the contracted quantum eigensolver (CQE), on quantum devices. On NISQ devices the addition of the $N$-representability conditions to the classical post-processing of the shadows will be useful not only in decreasing the number of measurements but also in mitigating the errors in the shadows during the reconstruction of the 2-RDM. The benefits of the enhanced shadow tomography---reduction in measurements and enhancement in accuracy---may be especially useful for resolving nearly degenerate excited states such as those arising in the study of conical intersections and non-adiabatic dynamics. Combining shadow tomography with $N$-representability conditions via semidefinite programming provides an important step towards the more realistic simulation of molecular excited-state phenomena on quantum devices.

\begin{acknowledgments}
D.A.M. gratefully acknowledges the U.S. National Science Foundation Grant CHE-2155082 and the Department of Energy, Office of Basic Energy Sciences, Grant DE-SC0019215. I.A. gratefully acknowledges the NSF Graduate Research Fellowship Program under Grant No. 2140001.
\end{acknowledgments}

\bibliography{references}

\end{document}